%% file: main_arxive.tex
\documentclass[aps,pre,twocolumn,floatfix,superscriptaddress]{revtex4-2}

\usepackage{graphicx}
\usepackage[english]{babel}
\usepackage{amsmath}
\usepackage{amssymb}
\usepackage{soul,color}

\begin{document}

\title{Acceleration of digital memcomputing by jumps}

\author{Yuriy~V.~Pershin}
\email{pershin@physics.sc.edu}
\affiliation{Department of Physics and Astronomy, University of South Carolina, Columbia, SC 29208 USA}

\begin{abstract}
In this article, we present the potential benefits of incorporating jumps into the dynamics of digital memcomputing machines (DMMs), which have been developed to address complex optimization problems. We illustrate the potential speed improvement of a DMM solver with jumps over an unmodified DMM solver by solving Boolean satisfiability (SAT) problems of different complicatedness. Our findings suggest that jumps can modify scaling exponents and improve solving times by up to 75~\%. Interestingly, the advantages of jumps can be seen in cases where the size of the jump is so large that otherwise the continuous dynamics of voltage variables becomes almost binary. 
\end{abstract}

\maketitle

\section{Introduction}

Digital memcomputing machines (DMMs)~\cite{Traversa17a,Sean3SAT,memcomputingbook} have been developed to solve factorization and combinatorial optimization problems efficiently. This form of computing~\cite{diventra13a} makes use of the continuous dynamics of intricate physical systems that encode the solution of problems of interest in their stable fixed points (if a solution or solutions exist). Importantly, the dynamics of such systems is guaranteed to reach a fixed point, if present, from any initial condition. 
Although it is hypothetically possible to build these machines with analog electronics~\cite{pershin23a,mem_analog_23a}, most current research focuses on software simulations of ordinary differential equations that describe these machines. Recently, an FPGA (field-programmable gate array) implementation of memcomputing equations was demonstrated~\cite{chung23}.

In this paper, we consider some specific DMMs~\cite{Sean3SAT} that were designed to tackle SAT problems. These machines rely on voltage variables and memory variables that are discussed, for example, in~\cite{Sean3SAT} and in the book on memcomputing~\cite{memcomputingbook}. The voltage variables serve as continuous counterparts to the binary variables in the problem under consideration, such as SAT. Meanwhile, memory variables are used to monitor resolved and unresolved relationships in the problem. Over time, the weights of unresolved relationships are increased, and their resolution is promoted.
In this regard, the operation of DMMs is similar to analogSAT~\cite{zoltan}. Compared to analogSAT~\cite{zoltan}, DMMs~\cite{Sean3SAT} use twice the number of auxiliary (memory) variables and different evolution functions. In addition, the adjustment of various parameters
can enhance the performance of DMMs.

Below, we explore a modified DMM approach that, from the point of view of variables, can be seen as a hybrid between those relying on continuous variables~\cite{zoltan,Sean3SAT} and those that rely on binary variables, such as~\cite{morris1993breakout,selman1993empirical}. The latter are local search algorithms that also incorporate clause weighting~\cite{morris1993breakout,selman1993empirical}. In these algorithms, a dynamic weight is associated with each clause (constraint)~\cite{morris1993breakout,selman1993empirical}. 
The weights of unsatisfied clauses are increased at each step and used as factors to decide which variable to flip next~\cite{morris1993breakout,selman1993empirical}. 

\begin{figure}[b]
\centering
\includegraphics[width=0.9\columnwidth]{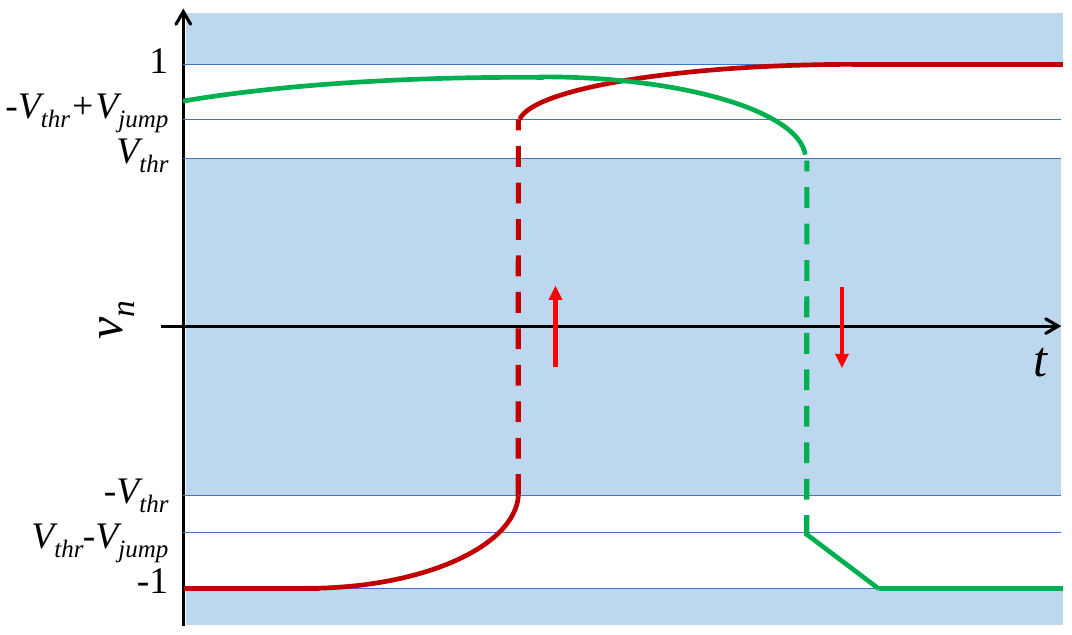}
\caption{Schematics of two voltage trajectories in a DMM with jumps.
 The shaded regions indicate the areas not allowed for the voltage variables.}
\label{fig:1}
\end{figure}

Fig.~\ref{fig:1} visually represents our main idea, showcasing the unique addition to the modified DMM algorithm.
Basically, whenever a voltage trajectory crosses a threshold voltage, $\pm V_{thr}$,
 the corresponding voltage variable changes by a jump voltage $\mp V_{jump}$, respectively.
Importantly, if the parameters $V_{thr}\geq 0$ and $V_{jump}>0$ meet the conditions $0<V_{thr}<1$ and $V_{thr}<V_{jump}/2$, the voltage variables become excluded from the interval $\{-V_{thr},V_{thr}\}$. Therefore, their evolution is restricted to two bands bounded by $\pm 1$ as it is shown schematically in Fig.~\ref{fig:1}.

The objective of this study is to understand the influence of jumps on the time-to-solution (TTS) of DMMs. We investigate the solution of 3-SAT problems with and without jumps by varying the jump size and threshold. We analyze both easy and difficult 3-SAT problems using conventional approaches for their generation~\cite{barthel2002,kowalsky20223}. In our simulations,  numerical integration is performed using forward Euler, which requires the least number of function evaluations compared to other integration schemes~\cite{zhang2021directed} and is advantageous for digital hardware implementations, such as~\cite{chung23}. Overall, we have observed that DMMs can be improved by jumps.

This paper is structured as follows. In Sec.~\ref{sec:2}, we present the equations, generation of instances, and simulations. Sec.~\ref{sec:3} contains the results of our numerical simulations. Sec.~\ref{sec:4} discusses the results. Finally, the paper concludes with conclusions.

\section{Model and simulations} \label{sec:2}

\subsection{Boolean satisfiability problems}

3-SAT is a Boolean satisfiability problem in which each clause is a disjunction of three Boolean variables or their negations. 
An example of a 3-SAT problem is 
\begin{equation}
    \nonumber (x_1\vee \bar{x}_2\vee x_3)\wedge (\bar{x}_1\vee \bar{x}_6\vee x_8)\wedge \cdots \;\;\;,
\end{equation}
where $x_i$ are Boolean variables, $\vee$ is OR, $\wedge$ is AND, and the bar denotes the negation. In the problem above, the first clause can be satisfied by
setting, e.g., $x_1$ to {\bf true} (1) or $x_2$ to {\bf false} (0). Another possibility is to assign {\bf true} to $x_3$. The goal is to find an assignment of variables that makes all clauses evaluate to true.

This study is based on three types of 3-SAT instances with planted solutions. The first type consists of the three-regular three-XORSAT instances. These instances were created as follows. First, a set of modulo 2 addition equations was generated by randomly selecting variables. Each equation consisted of three variables, with their negation randomly assigned. Moreover, each variable was ensured to appear in exactly three equations. The variables were then randomly assigned, and, subsequently, each equation was converted to four 3-SAT clauses.

Two other types of instances were created following the approach of Barthel et al.~\cite{barthel2002} at $p_0=0.08$ using $M/N=4.3$ and $M/N=7$, where $M$ is the number of clauses and $N$ is the number of variables. For more details, we refer to Barthel et al.~\cite{barthel2002}.

According to previous studies, DMMs exhibit a polynomial scaling for Barthel instances~\cite{Sean3SAT}, whereas for XORSAT instances, the scaling is exponential~\cite{kowalsky20223}. The latter result is anticipated to be improved by selecting more suitable parameters~\footnote{M. Di Ventra, private communication}. Regardless, in this article, XORSAT instances are refferred to as very difficult, Barthel $M/N=4.3$ as difficult, and Barthel $M/N=7$ as easy.

\subsection{DMMs with jumps}

The model considered in this work is a DMM 3-SAT solver~\cite{Sean3SAT} improved by adding jumps.
For a 3-SAT problem of $N$ variables and $M$ clauses, the basic DMM equations are written as
\begin{eqnarray}
\dot{v}_n&=&\sum\limits_mx_{l,m}x_{s,m}G_{n,m}(v_n,v_j,v_k)+\left( 1+\zeta x_{l,m}\right)\cdot \nonumber \\
& & \left(1-x_{s,m}\right) R_{n,m}(v_n,v_m,v_k), \label{eq:mc1}\\
\dot{x}_{s,m}&=&\beta \left( x_{s,m} +\epsilon \right)\left( C_m(v_i,v_j,v_k)-\gamma\right), \label{eq:mc2}\\
\dot{x}_{l,m}&=&\alpha \left( C_m(v_i,v_j,v_k)-\delta\right), \label{eq:mc3}\\
G_{n,m}&=&\frac{1}{2}q_{n,m}\text{min}\left[\left( 1-q_{j,m}v_j\right), \left( 1-q_{k,m}v_k\right)\right], \label{eq:mc4}\\
R_{n,m}&=&\begin{cases}
    \frac{1}{2}\left(q_{n,m}-v_n \right), \\
    \hspace{1cm} \text{if } C_m(v_n,v_j,v_k)=\frac{1}{2}\left(1-q_{n,m}v_n \right),\;\;\\
    0, \hspace{7mm} \text{otherwise}.
  \end{cases} \label{eq:mc5}
\end{eqnarray}
Here, $v_n$ are the continuous voltage variables ($n=1,\ldots,N$), $x_{s,m}$ and $x_{l,m}$ are the memory variables (where $s$ stands for short, $l$ stands for long, and $m=1,\ldots,M$), and $q_{j,m}=\{-1,0,1\}$ describes the contribution of $j$-s voltage variable to clause $m$. Moreover, $\alpha$, $\beta$, $\gamma$, $\delta$, $\epsilon$ and $\zeta$ are constants~\cite{Sean3SAT}. 
Furthermore, $v_n$-s are restricted to the interval $[-1,1]$, $x_{s,m}$-s are restricted to the interval $[\epsilon, 1-\epsilon]$, and $x_{l,m}$-s to the interval $[1,10^4M]$.

The first term in Eq.~(\ref{eq:mc1}) can be interpreted as a ``gradient-like'' term, while the second -- as a ``rigidity'' term~\cite{Sean3SAT}. The purpose of the ``rigidity'' term is to suppress the evolution of $v_n$ when its value is the best to satisfy clause $m$. The clause function $C_m(v_i,v_j,v_k)$ is defined as
\begin{align}
& C_m(v_i,v_j,v_k)= \nonumber \\     
 &  \hspace{3mm}  \frac{1}{2}\text{min}\left[\left(1-q_{i,m}v_i \right),\left(1-q_{j,m}v_j \right),\left(1-q_{k,m}v_k \right)\right].\label{eq:6}
\end{align}
This function characterizes the state of the variable that most closely satisfies the clause $m$. More information on the basic DMM model can be found in Refs.~\cite{Sean3SAT,memcomputingbook}.

\begin{figure*}[bt]
\centering
\includegraphics[width=0.32\textwidth]{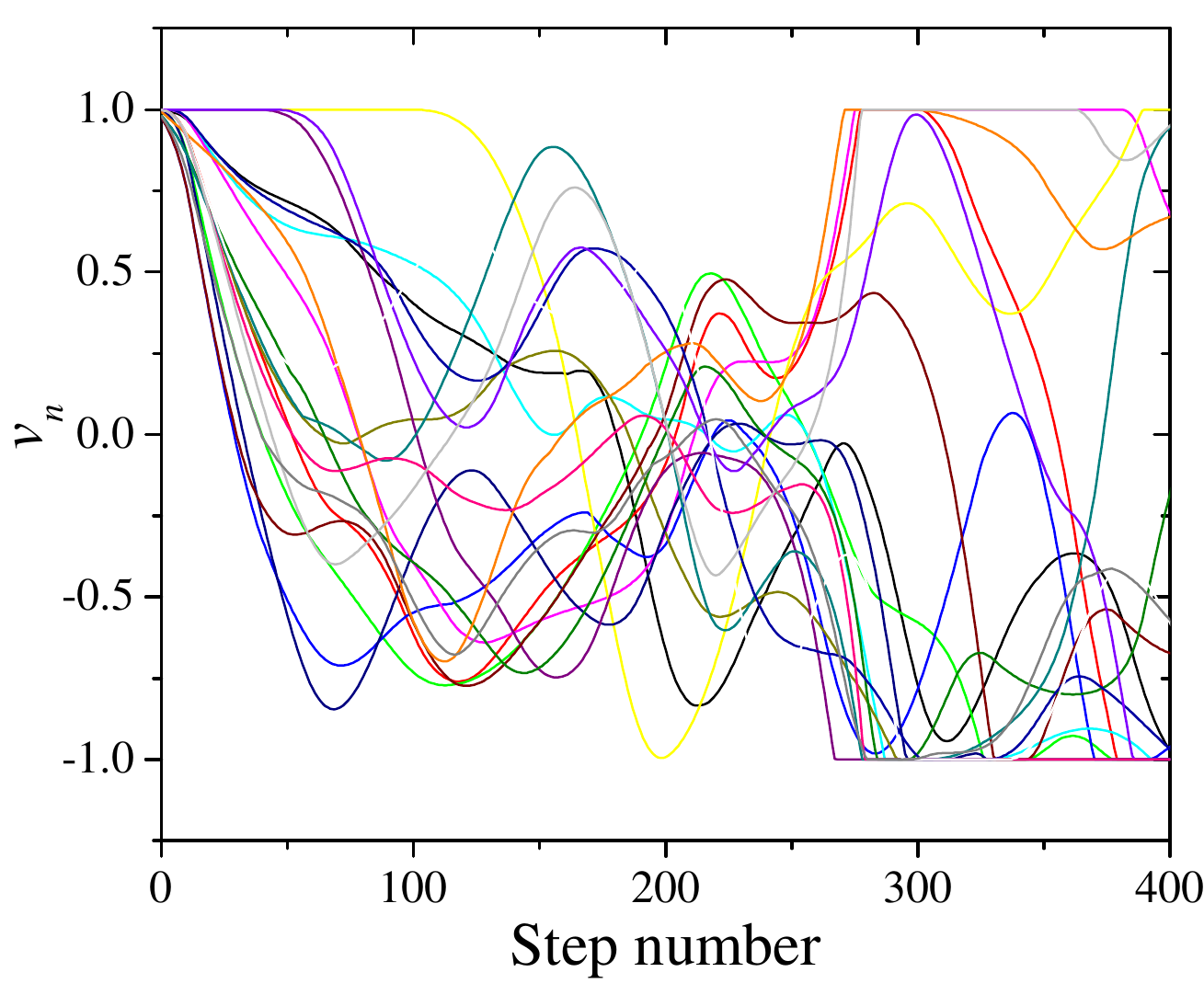}
\includegraphics[width=0.32\textwidth]{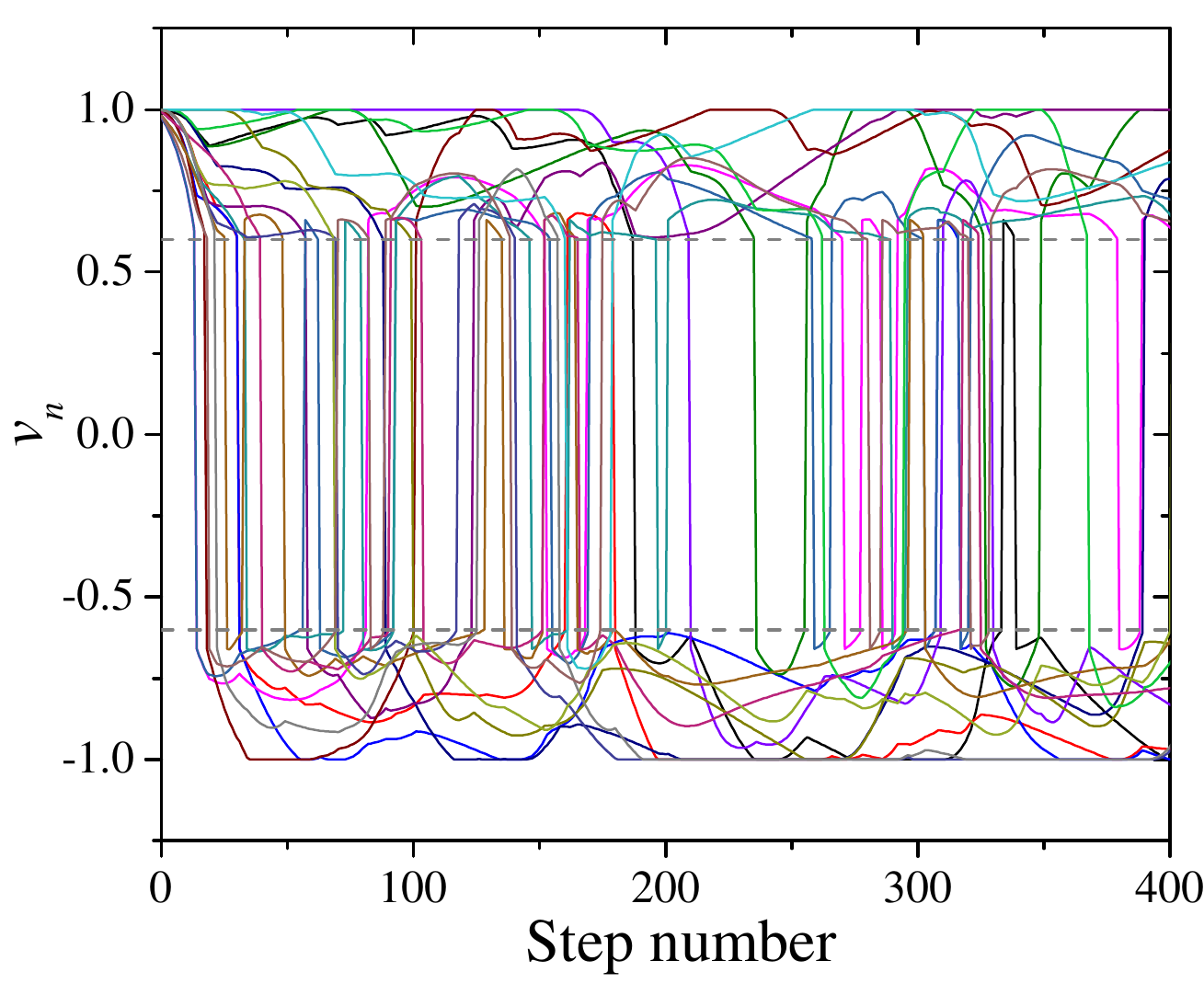}
\includegraphics[width=0.32\textwidth]{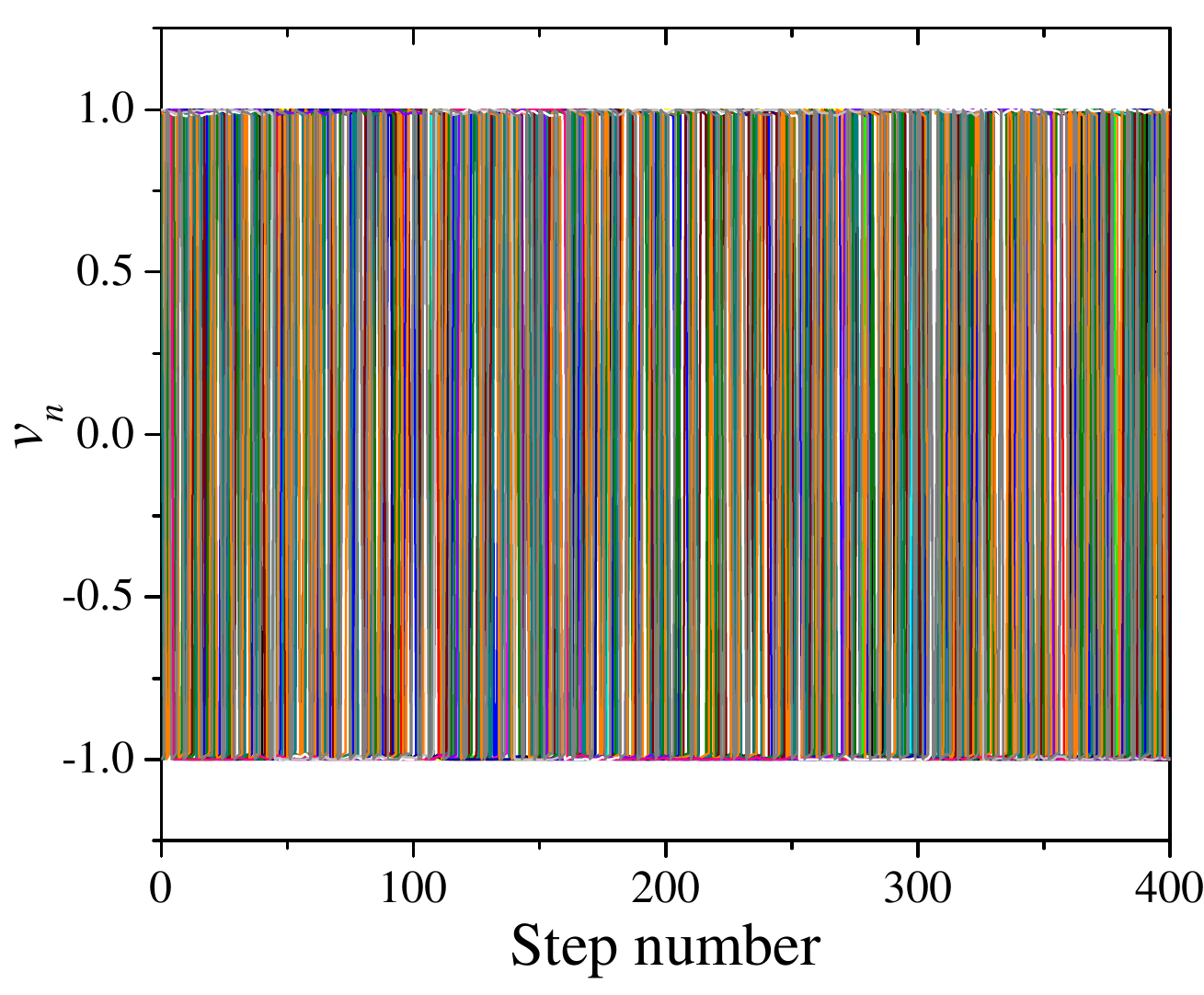}
(a) \hspace{5cm} (b) \hspace{5cm} (c)
\caption{Voltage trajectories in (a) unmodified DMM and (b), (c) DMM with jumps. The plots were generated using $N=20$ XORSAT instances and deterministic initial conditions $v_n(t=0)=1$, where $n=1,..,20$. Furthermore, in (b), we set $V_{thr}=0.6$ and $V_{jump}=2.1V_{thr}$, and in (c), we set $V_{thr}=0.98$ and $V_{jump}=2.1V_{thr}$.}
\label{fig:2}
\end{figure*}

The jumps were integrated into the dynamics of the voltage variables $v_n$. To achieve this,  the voltage variables are compared with the thresholds $\pm V_{thr}$ at each integration step. Whenever a voltage trajectory crosses $\pm V_{thr}$, the corresponding variable is changed to
$\pm(V_{thr}-V_{jump})$ if $|V_{thr}-V_{jump}|<1$ or $\mp 1$ otherwise, respectively. Fig.~\ref{fig:1} presents a schematic illustration of jumps in a DMM with jumps.

\subsection{Numerical simulations}

Numerical simulations were performed with the forward Euler method using the following set of parameter values: $\alpha=5$, $\beta=20$, $\gamma=1/4$, $\delta=1/20$, $\epsilon=0.1$, $\zeta=0.1$.
In our simulations, we utilized different values for the threshold voltage and jump voltage that are provided below for each specific simulation. 
Unless otherwise stated, the voltage variables were initially randomly assigned using a uniform random distribution in the interval $[-1,1]$.
Memory variables were initialized as follows: $x_{s,m}(0)=\epsilon$ and $x_{l,m}(0)=1$.

The C++ programming language was used to implement the dynamic equations (\ref{eq:mc1})-(\ref{eq:mc3}). Numerical integration was carried out with a time step of $\Delta t=0.01$.
The simulations were stopped if the solution could not be found within a specified number of steps.
Typically, we conducted $10^3$ or $10^4$ simulations for each set of parameters and instance type. These simulations were utilized to determine the median TTS and generate distributions of TTS. We emphasize that the times reported in this work are the intrinsic times of DMM dynamics. These should not be confused with the simulation times on conventional CPUs.


\section{Results} \label{sec:3}

\subsection{Trajectories}
Fig.~\ref{fig:2} presents examples of voltage trajectories in the unmodified DMM (Fig.~\ref{fig:2}(a)) and modified DMM that incorporates jumps (Fig.~\ref{fig:2}(b) and (c)). One can observe that the presence of jumps has greatly changed the dynamics of DMM. Fig.~\ref{fig:2}(b) and (c) show that on average the frequency of jumps increases with the threshold voltage $V_{thr}$. This finding aligns with expectations as a higher threshold voltage results in a reduced phase space for the system's dynamics. Consequently, in this scenario, the trajectories encounter boundaries more often.

In particular, Fig.~\ref{fig:2}(c) shows the trajectories in a DMM with a significantly reduced phase space. In fact, at $V_{thr}=0.98$ the phase space for a single voltage variable is just 2~\% of that in the unmodified DMM, while for all voltage variables the phase space is reduced by a factor of $1/(0.02)^N\approx 10^{34}$. According to Fig.~\ref{fig:2}(c), in this case, jumps occur almost continuously, so that a jump or multiple jumps occur at nearly every integration step. In the following, we investigate the impact of such frequent jumps (and less frequent jumps) on the time required to reach a solution.

\subsection{Time-to-solution}

\begin{figure}[b]
\centering
(a) \;\includegraphics[width=0.7\columnwidth]{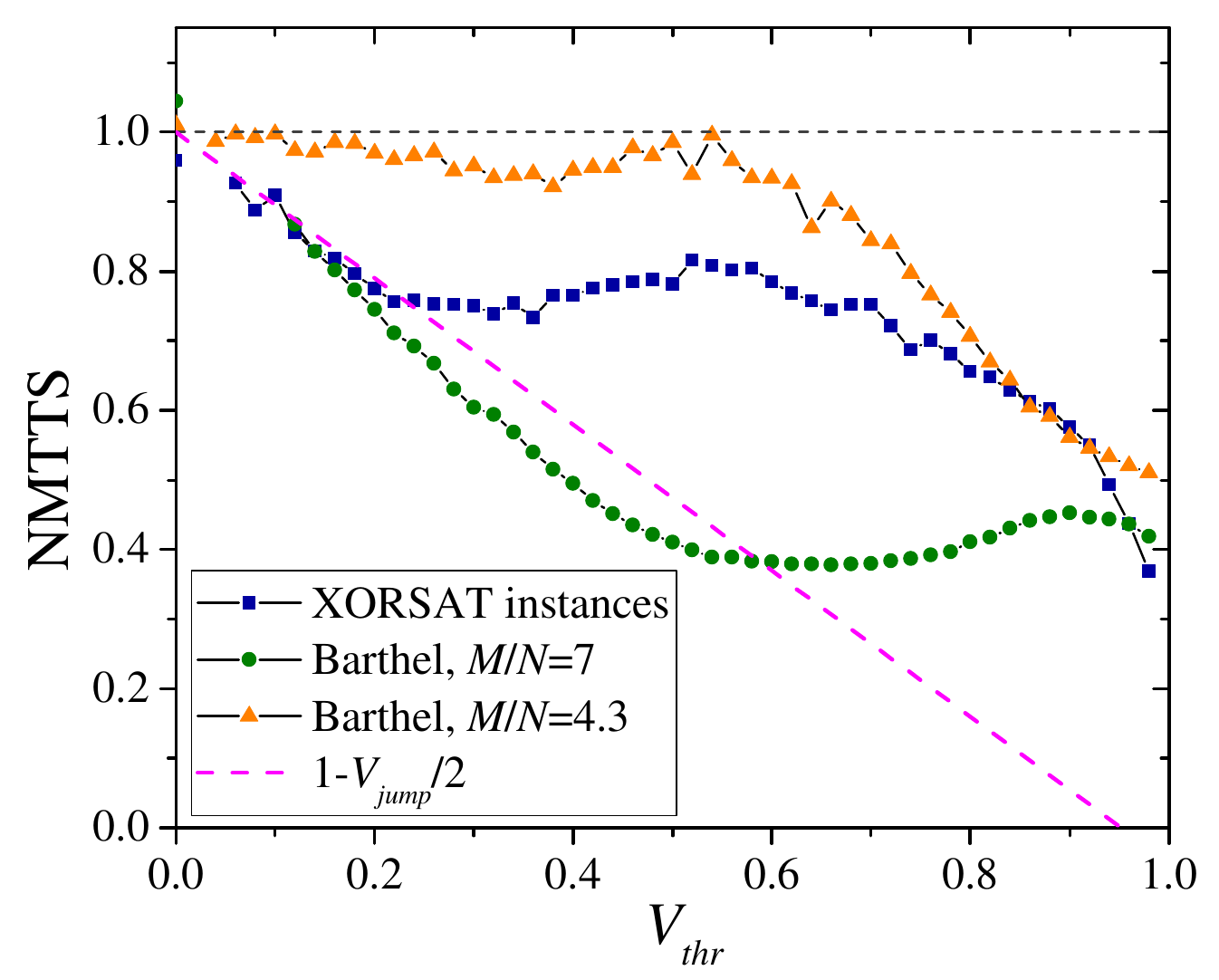}\\
(b) \;\includegraphics[width=0.7\columnwidth]{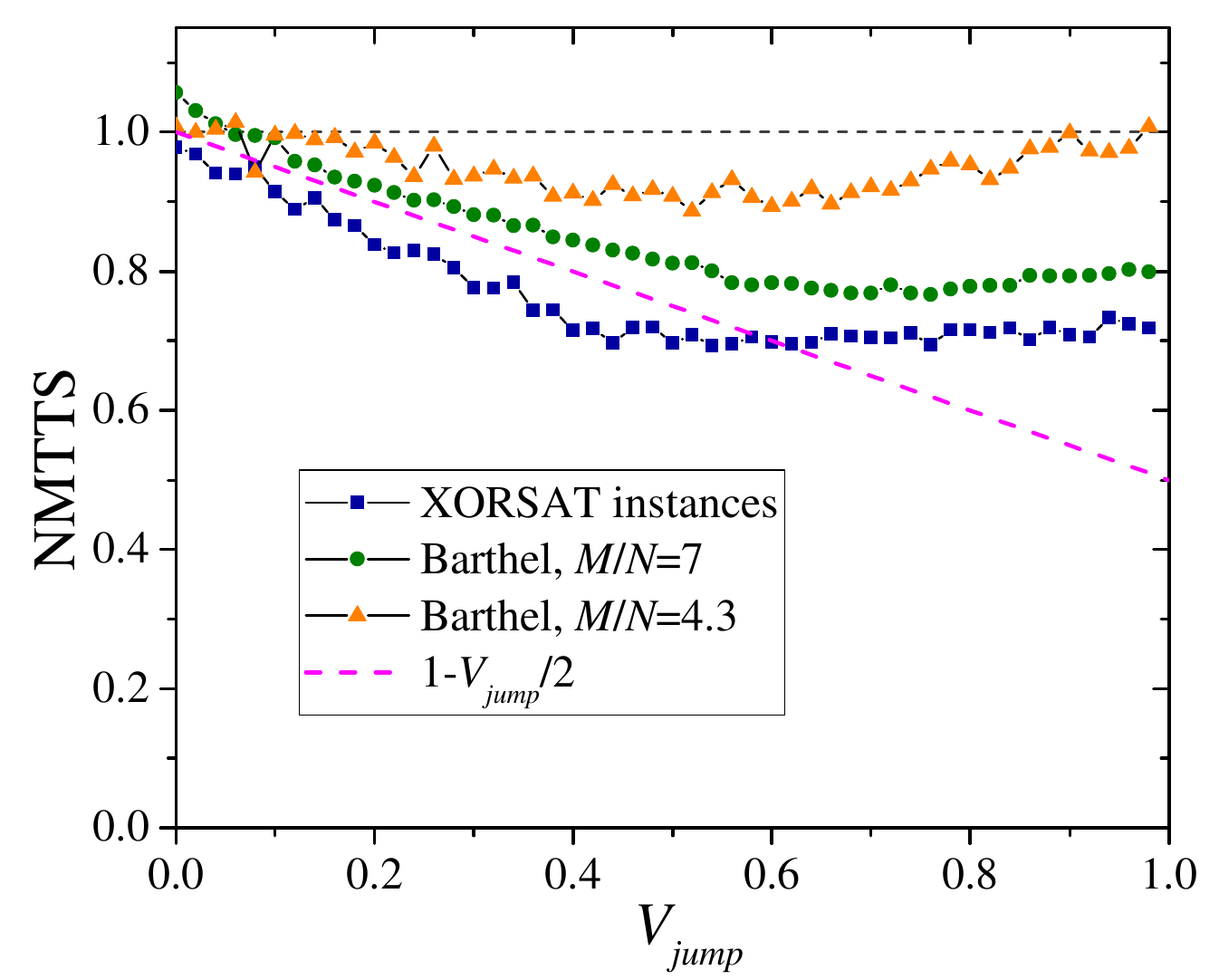}
\caption{The normalized median time-to-solution (NMTTS) in a DMM with jumps, relative to the time-to-solution in the unmodified DMM for (a) $V_{jump}=2.1V_{thr}$ and (b) $V_{thr}=0$. To generate these graphs, we used $N=50$ XORSAT instances, $N=1000$ Barthel instances at $M/N=7$, and $N=100$ Barthel instances at $M/N=4.3$. The dashed lines are model curves discussed in the text.}
\label{fig:3}
\end{figure}

\begin{figure*}[tb]
\centering
\includegraphics[width=0.32\textwidth]{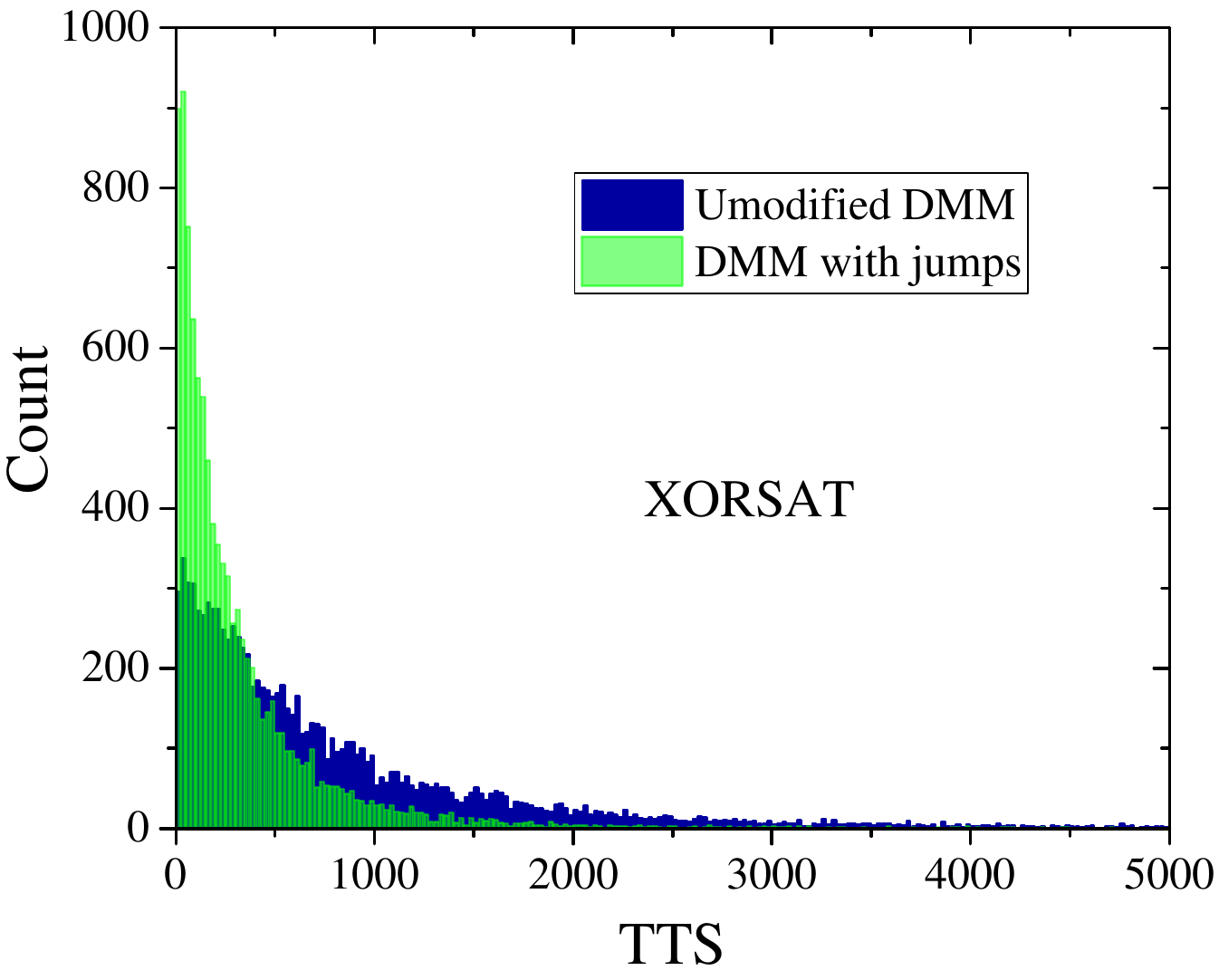}
\includegraphics[width=0.32\textwidth]{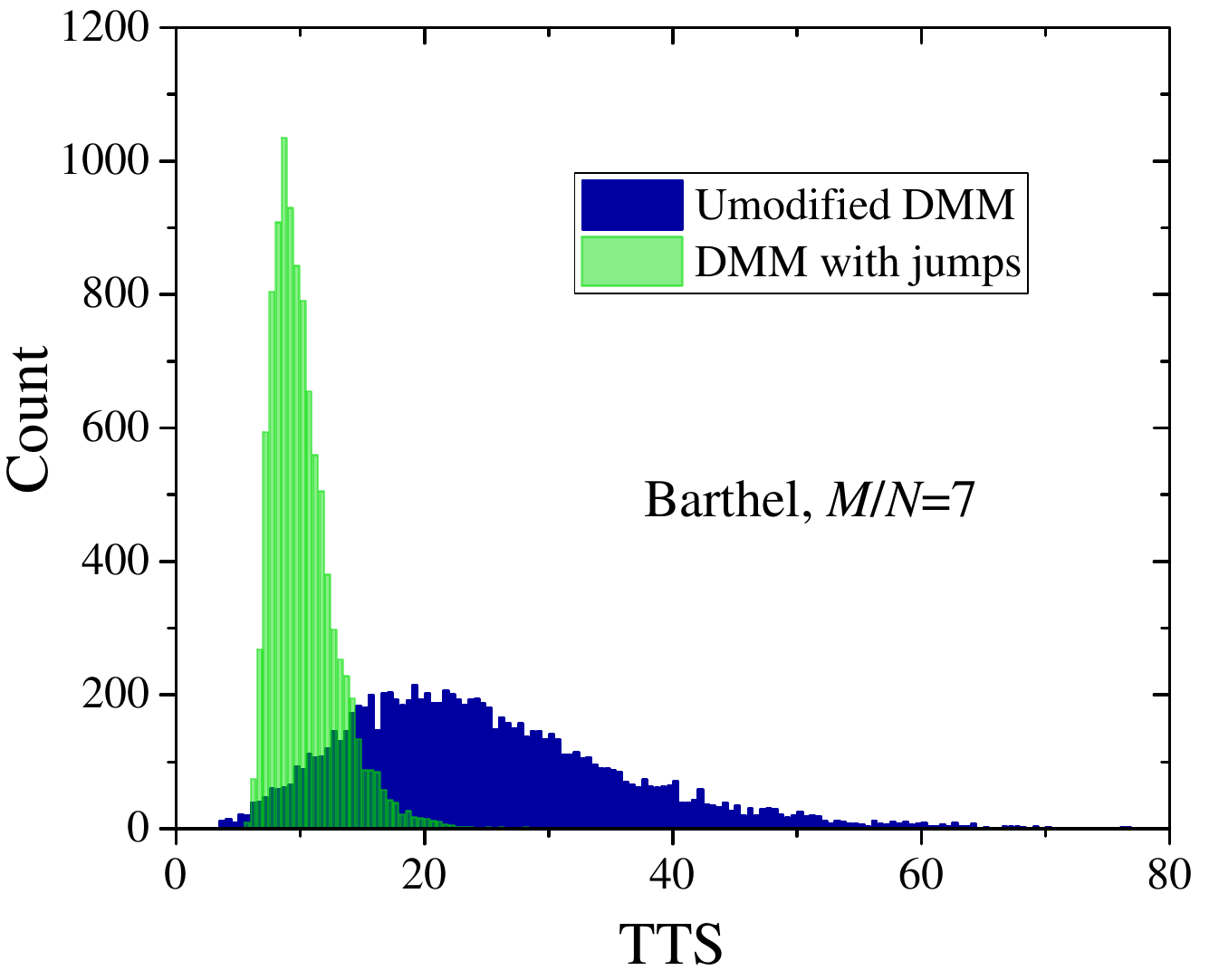}
\includegraphics[width=0.32\textwidth]{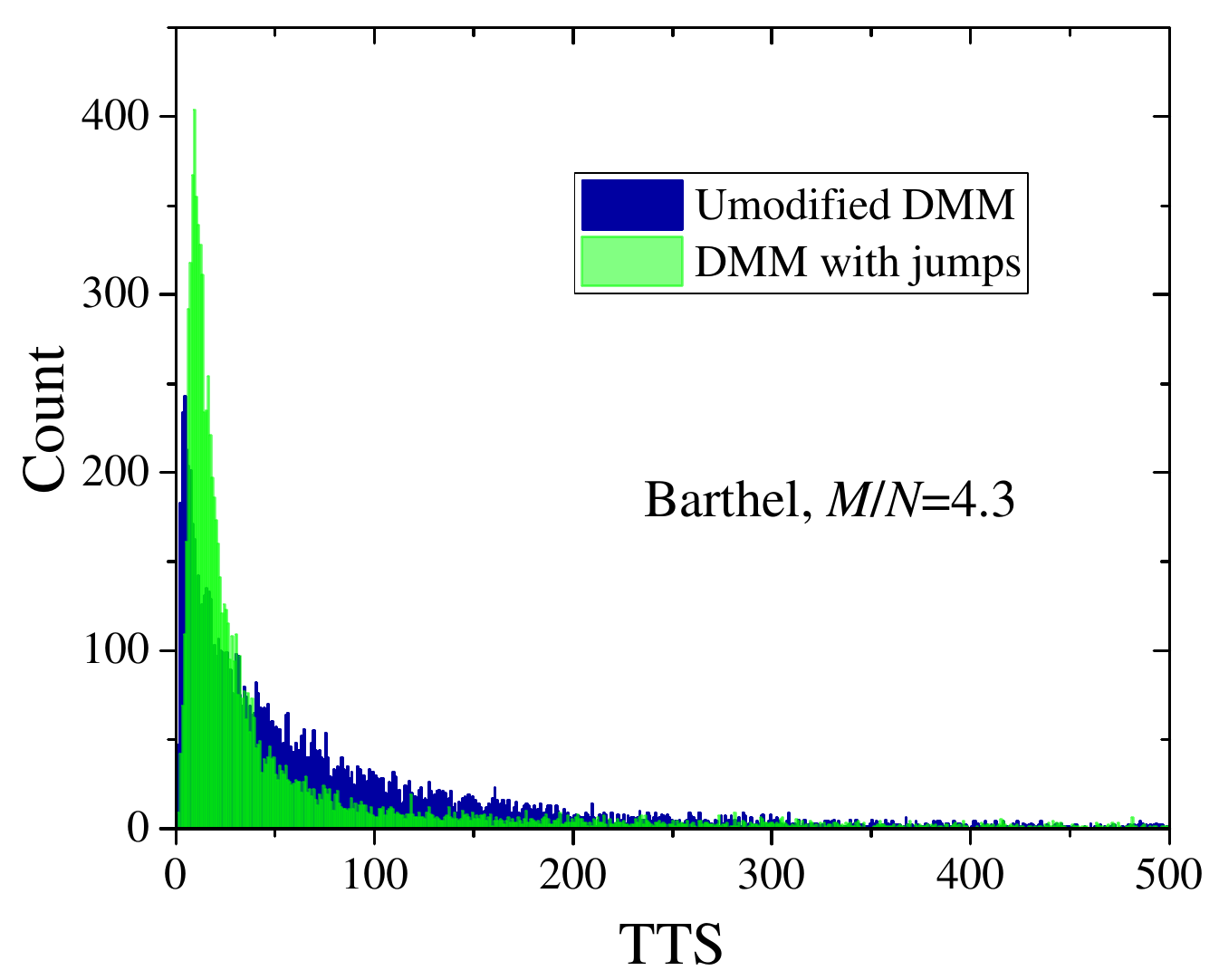}
(a) \hspace{5cm} (b) \hspace{5cm} (c)
\caption{Distributions of TTS found with the unmodified DMM and DMM with jumps for (a) XORSAT instances, (b) Barthel $M/N=7$ instances, and (c) Barthel $M/N=4.3$ instances. The distributions for DMM with jumps (green color) correspond to $V_{thr}=0.98$ in Fig.~\ref{fig:3}(a) (the rightmost points in Fig.~\ref{fig:3}(a)). The size of the bin is $w=25$ in (a), $w=0.5$ in (b), and $w=1$ in (c).}
\label{fig:4}
\end{figure*}

The median TTS is a widely used benchmark to evaluate the performance of various solvers. Fig.~\ref{fig:3} shows the median time-to-solution in a DMM with jumps normalized by the time-to-solution in the unmodified DMM for the three types of 3-SAT instances: XORSAT, Barthel $M/N=7$, and Barthel $M/N=4.3$. 
Each point in Fig.~\ref{fig:3} was obtained by solving $10^4$ instances of SAT. We mention that there are certain initial points (at $V_{thr}\leq 0.1$) in Fig.~\ref{fig:3}(a) that are not included because (at corresponding values of $V_{thr}$) the solution was not obtained within a specified number of integration steps for the majority of problem instances. 

The findings presented in Fig.~\ref{fig:3} suggest that the effect of jumps depends on the type of instances, $V_{thr}$, and $V_{jump}$. Fig.~\ref{fig:3}(a) shows the normalized median TTS 
for the case when the phase space of voltage variables is reduced to two bands (as shown schematically in Fig.~\ref{fig:1}). As per Fig.~\ref{fig:3}(a), the most significant effect is observed for easy problems, followed by the intermediate effect for very difficult problems, and the smallest effect for difficult problems. We also note that the rightmost points in Fig.~\ref{fig:3} correspond to $U_{thr}=0.98$, and the step of $U_{thr}$ in Fig.~\ref{fig:3} is $0.2$. In a smaller step calculations in the interval $[0.99,1]$ we have observed a sharp increase of TTS as $V_{thr}\rightarrow 1$.

Fig.~\ref{fig:3}(b) illustrates the normalized median TTS as a function of $V_{jump}$ at $V_{thr}=0$. In this scenario, the voltage variables occupy the original domain $[-1,1]$. One can observe that in Fig.~\ref{fig:3}(b), the effect of jumps is, in principle, comparable to that in Fig.~\ref{fig:3}(a) (taking into account that the largest jump in Fig.~\ref{fig:3}(b) is one half of that in Fig.~\ref{fig:3}(a)). Moreover, in Fig.~\ref{fig:3}(b) the order of the curves is different: The most significant effect is now observed for very difficult problems, followed by the intermediate effect for easy problems, and the smallest effect for difficult problems.

Examples of TTS distributions for very difficult, easy, and difficult problems are shown in Fig.~\ref{fig:4}(a)-(c), respectively. In this figure, we are comparing the distributions obtained using the original DMM and DMM with jumps. In all cases, we observe that the jumps narrow the TTS distributions and shift the distributions to shorter times. 

In the Supplementary Material (SI), we employed an exponential function and an inverse Gaussian function to fit the distributions shown in Fig.~\ref{fig:4} (see Figs. S1-S3 in the SI). In the case of very difficult problems shown in Fig.~\ref{fig:4}(a), the TTS distributions in the unmodified DMM and modified DMM can be fitted by exponential functions (see Fig.~S1 in the SI). 
In the case of easy problems shown in Fig.~\ref{fig:4}(b), the TTS distributions for the unmodified DMM and the modified DMM can be fitted by inverse Gaussian distributions (see Fig.~S2 in the SI). In the case of difficult problems shown in Fig.~\ref{fig:4}(c), the TTS distributions cannot be well fitted by the exponential function or the inverse Gaussian function (see Fig.~S3 in the SI).

\subsection{Scaling}

\begin{figure}[bt]
\centering
\includegraphics[width=0.7\columnwidth]{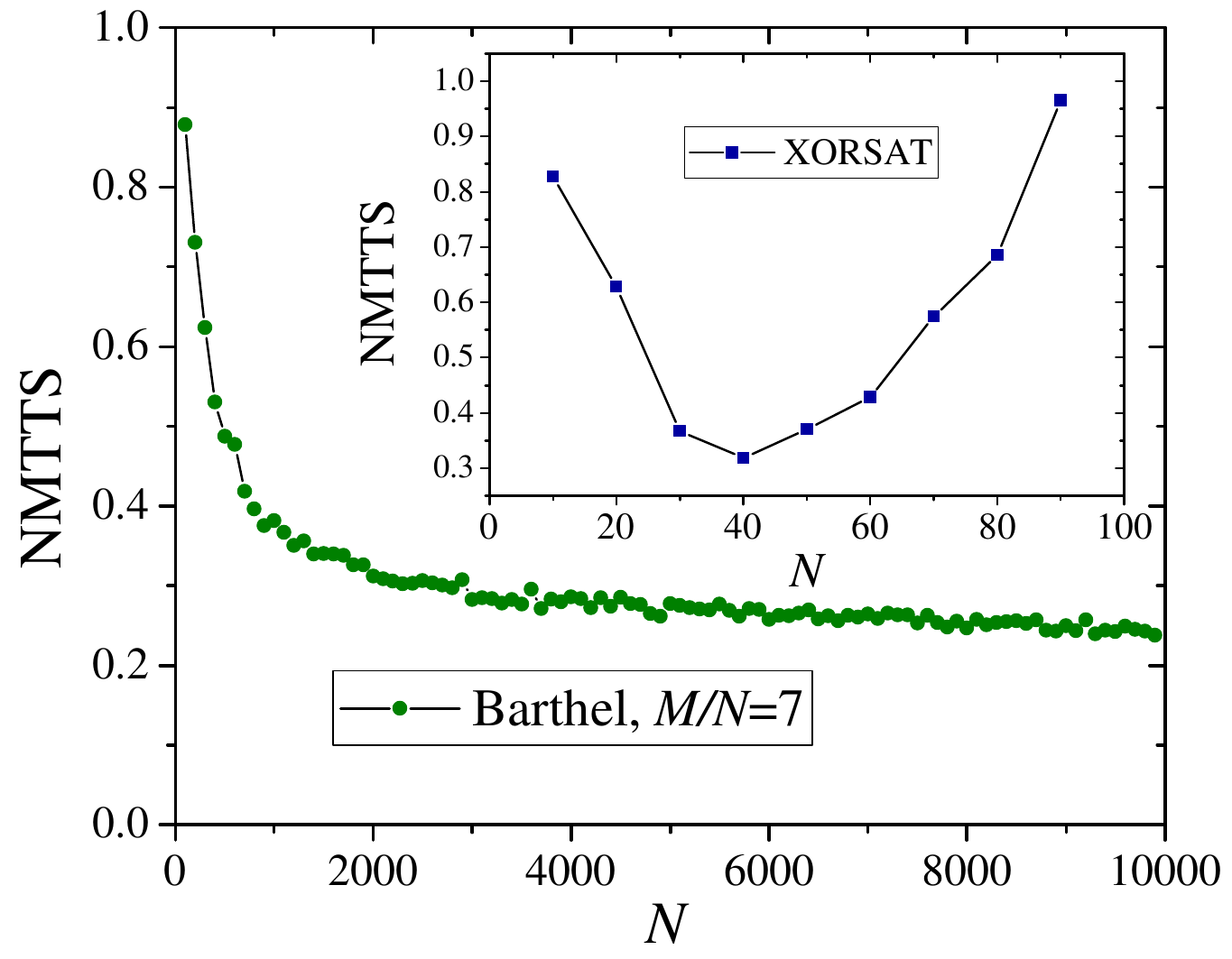}
\caption{The normalized median TTS as a function of the problem size found with the DMM with jumps, relative to TTS in the unmodified DMM for the same problem size. The main plot was obtained using easy instances, $V_{thr}=0.65$, and $V_{jump}=2.1V_{thr}$. The inset was obtained using very difficult instances, $V_{thr}=0.98$, and $V_{jump}=2.1V_{thr}$.}
\label{fig:5}
\end{figure}

In some of our simulations, we have changed the size of the problem (specifically, the number of variables $N$) while keeping all other parameters constant. Fig.~\ref{fig:5} presents the results of these simulations obtained for easy and very difficult problems. Two main observations can be drawn from Fig.~\ref{fig:5}. First, for easy problems, the normalized median TTS decreases with the size of the problem. Second, in the case of very difficult problems, the normalized median TTS reaches a minimum at approximately $N= 40$, increasing as $N$ continues to increase. Certainly, these findings are constrained to the parameter values employed in our study.

The TTS data used to generate Fig.~\ref{fig:5} are shown without normalization in Fig.~S4 in the SI. It is interesting to note that jumps alter the scaling exponents for both simple and very difficult problems. A significant improvement is evident for simple problems, with the scaling exponent shifting from approximately 0.43 to around 0.25, representing a noteworthy speed-up.

\section{Discussion}  \label{sec:4}

In this paper, we have presented some initial simulations of DMMs with jumps. It has been observed that jumps have the potential to accelerate DMMs, and in some cases, the acceleration is significant. Currently, there is no theory available to explain the effect of jumps on TTS in DMMs. In what follows, we provide a simple analysis of this phenomenon.

The analysis is based on the straightforward observation that jumps speed-up the evolution of voltage variables. To quantify this effect, let us use a linear approximation for the change of voltage variables in the unmodified DMMs (when these variables switch from one logic state to another). The suitability of this approximation is supported by Fig.~\ref{fig:2}(a) showing very similar slopes in the dynamics of many voltage variables.

Assuming that $\Delta v_n = \alpha \Delta t$, where $\alpha$ is the rate, the switching times (for the switching from $-1$ to $1$ or vice versa) in the unmodified DMM and DMM with jumps can be approximated by $t_{sw}^{unmod}=2/|\alpha |$ and $t_{sw}^{mod}=(2-V_{jump})/|\alpha |$, respectively. In these expressions, the denominator represents $\Delta v_n$ to be traveled through by continuous dynamics. The ratio of these times,
\begin{equation}
    \frac{t_{sw}^{mod}}{t_{sw}^{unmod}}=1-\frac{V_{jump}}{2}, \label{eq:7}
\end{equation}
is the theoretical acceleration factor. Eq.~(\ref{eq:7}) is shown in Fig.~\ref{fig:3}(a) and (b) by dashed lines. 

We note that in Fig.~\ref{fig:3}(a), Eq.~(\ref{eq:7}) is in a reasonable agreement with our simulation results for Barthel $M/N=7$ instances for $V_{thr}\lesssim 0.5$ and XORSAT instances for $V_{thr}\lesssim 0.2$. Regarding Fig.~\ref{fig:3}(b), a reasonable agreement is observed in the case of the XORSAT and Barthel $M/N=7$ instances for $V_{jump}\lesssim 0.6$. 
Since the acceleration for Barthel $M/N=7$ instances in Fig.~\ref{fig:3}(a) and XORSAT instances in Fig.~\ref{fig:3}(b) is greater compared to the prediction of Eq.~(\ref{eq:7}), it 
may exist a different factor contributing to the acceleration.

Finally, we provide a brief discussion on the shape of the TTS distributions (Fig.~\ref{fig:4} and Figs. S1-S3 in the SI). According to a recent study by Primosch et al.~\cite{Primosch23a}, often, the polynomial algorithms lead to inverse Gaussian distributions, while exponential algorithms lead to exponential distributions. Our findings for easy and very difficult instances align with the conclusion in ~\cite{Primosch23a}. In the case of difficult instances, we have observed that the TTS distributions cannot be fitted well with either exponential function or inverse Gaussian function (see Fig.~S3 in the SI). It is important to emphasize again that there might be more suitable DMM parameters that could be utilized.

\section*{Conclusion}

In this study, we introduce DMMs with jumps, an enhanced method to solve intricate optimization problems. By making minor adjustments to the original DMM algorithm, we have successfully reduced the time-to-solution by up to 75~\% and modified scaling exponents. These findings are expected to have a positive impact on the development of more advanced digital memcomputing machines, including their hardware implementation.

\section*{Acknowledgements}
The author is thankful to M. Di Ventra, D. C. Nguyen, and Y.-H. Zhang for helpful comments. This work was supported by the National
Science Foundation grant number ECCS-2229880. 

\bibliographystyle{apsrev4-2}
\bibliography{Xbibli}

\newpage

\onecolumngrid
\setcounter{section}{0}
\setcounter{figure}{0}
\setcounter{equation}{0}

\include{SI_arxive}

\end{document}

%% file: SI_arxive.tex
\vspace{0cm}
\begin{center}
    {\LARGE \it Supplementary Material}    
\end{center}
\vspace{0.6cm}

\renewcommand{\thefigure}{S\arabic{figure}}
\renewcommand{\thesection}{S\arabic{section}}
\renewcommand{\theequation}{S\arabic{equation}}

\section{Fitting Distributions of TTS}

The time-to-solution (TTS) distributions were fitted using an exponential function and/or inverse Gaussian function chosen on the basis of visual examination of the calculated distributions of TTS. In all cases, the curves were normalized to the total number of instances, $N_{inst}$, used in each set of calculations. 

For the exponential function  we used
\begin{equation}
    f_{exp}(x)=\frac{N_{inst} w}{a} e^{-\frac{x}{a}},
    \label{eq:exp}
\end{equation}
where $w$ is the bin size and $a$ is a fitting constant. For the inverse Gaussian function we used
\begin{equation}
    f_{iG}(x)=N_{inst} w \sqrt{\frac{b}{2\pi x^3}}e^{-\frac{b(x-a)^2}{2a^2x}},
    \label{eq:iG}
\end{equation}
where $a$ and $b$ are fitting constants. The parameters of the fitting functions, Eqs.~(\ref{eq:exp}) and (\ref{eq:iG}), were determined using Wolfram Mathematica version 14.0.0.

\begin{figure}[hbt!]
(a) \includegraphics[width=0.45\columnwidth]{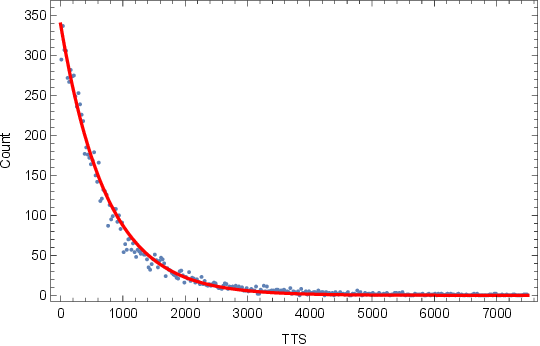} (b)
\includegraphics[width=0.45\columnwidth]{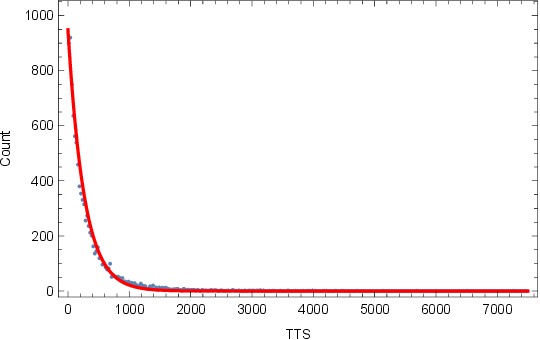}
\caption{Fitting distributions of TTS for (a) the unmodified DMM and (b) DMM with jumps for the case of XORSAT instances. The red lines represent Eq.~(\ref{eq:exp}) with $a=736.748\pm 5.39629$ in (a) and $a=263.52\pm 2.20221$ in (b). The data points are the same as in Fig.~4(a) (main text).}
\label{fig:1}
\end{figure}

\begin{figure}[h]
\centering
(a) \includegraphics[width=0.45\columnwidth]{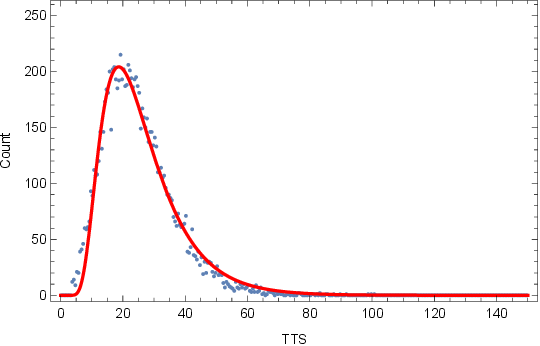} (b)
\includegraphics[width=0.45\columnwidth]{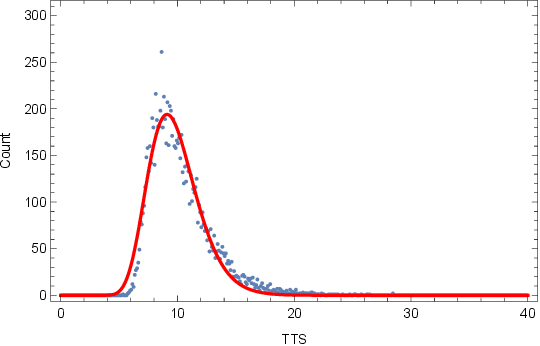}
\caption{Fitting distributions of TTS for (a) the unmodified DMM and (b) DMM with jumps for the case of Barthel $M/N=7$ instances. The red lines represent Eq.~(\ref{eq:iG}) with $a=26.269\pm 0.126056$, $b=113.28 \pm 1.55843$ in (a) and $a=9.81381 \pm 0.0265121$, $b=200.076 \pm 3.83326$ in (b).
The data points are the same as in Fig.~4(b) (main text), up to the bin size.}
\label{fig:2}
\end{figure}

\begin{figure}[h]
\centering
(a) \includegraphics[width=0.45\columnwidth]{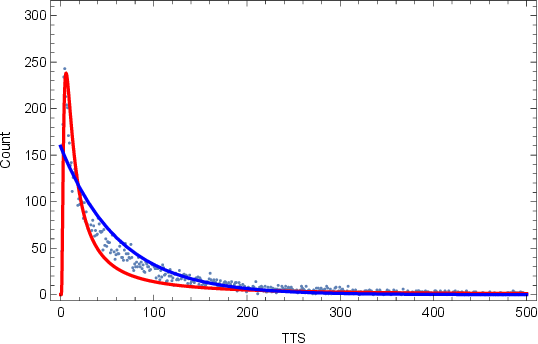} (b)
\includegraphics[width=0.45\columnwidth]{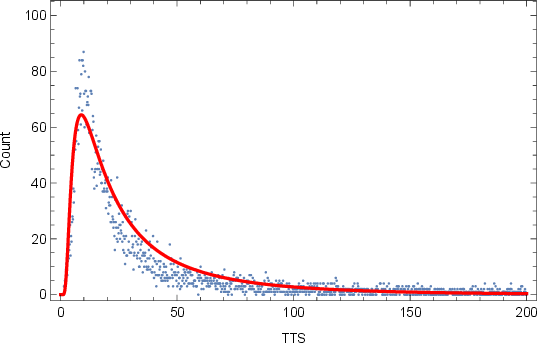}
\caption{Fitting distributions of TTS for (a) the unmodified DMM and (b) DMM with jumps for the case of Barthel $M/N=4.3$ instances. The red lines represent Eq.~(\ref{eq:iG}) with $a=-219.26\pm 39.7956$, $b=17.8928 \pm 0.312861$ in (a) and $a=36.4246\pm 0.676755$, $b=28.3518 \pm 0.409163$ in (b). The blue line in (a) represents Eq.~(\ref{eq:exp}) with $a=62.6837\pm 1.32475$. The data points are the same as in Fig.~4(c) (main text), up to the bin size.}
\label{fig:3}
\end{figure}

\clearpage

\section{Scalability}

\begin{figure}[hbt!]
\centering
(a) \includegraphics[width=0.45\columnwidth]{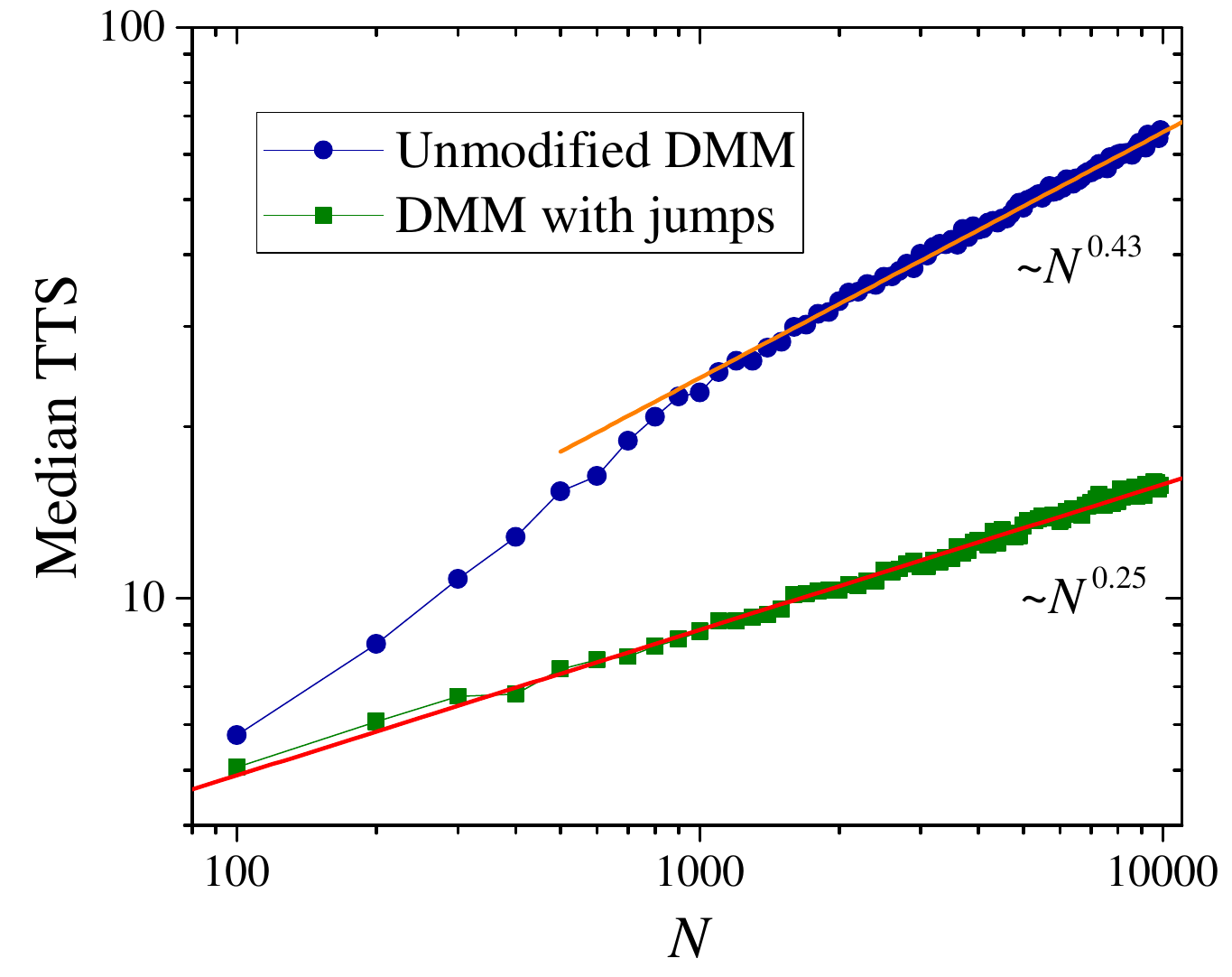} (b)
\includegraphics[width=0.45\columnwidth]{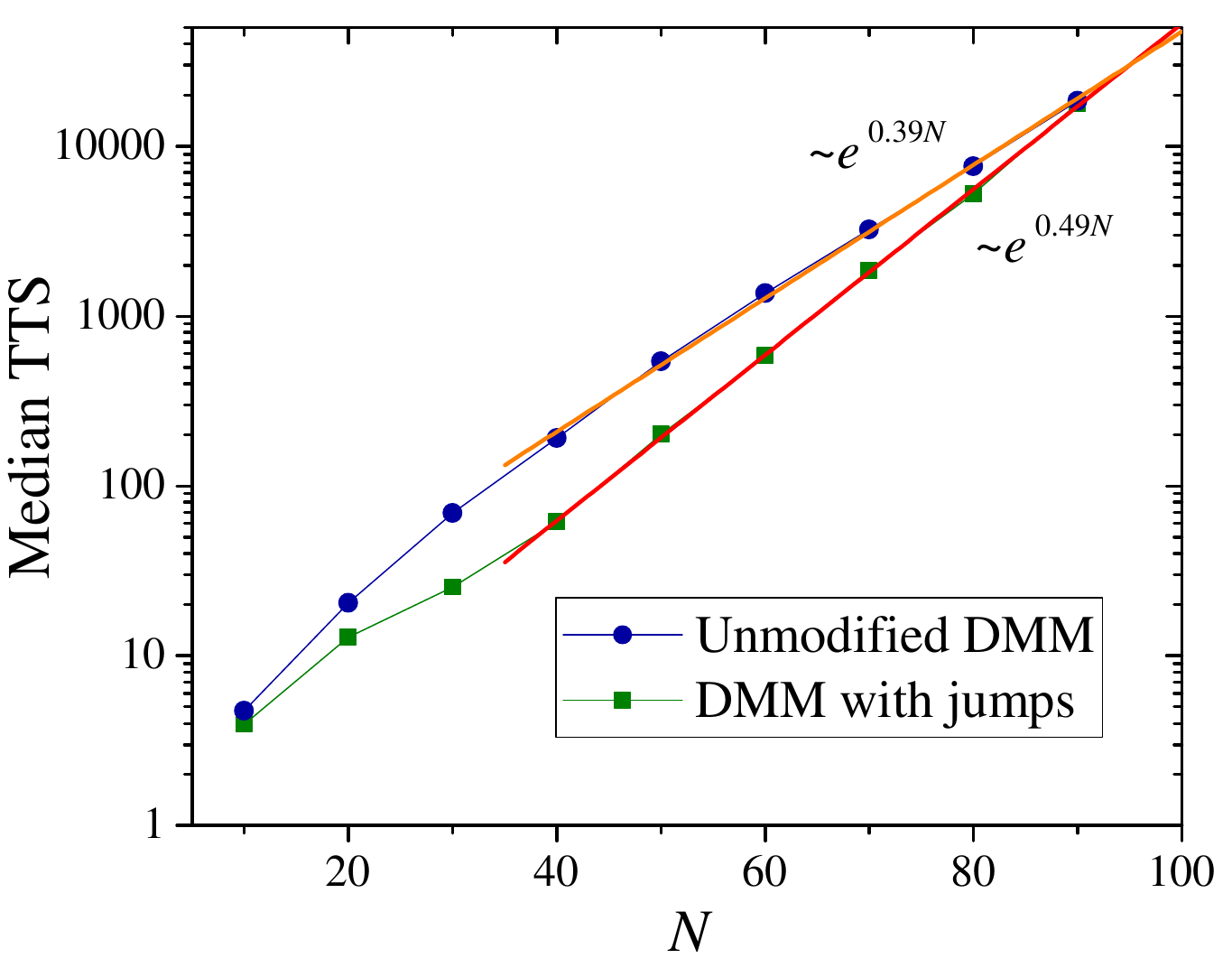}
\caption{Scalability of (a) Barthel $M/N=7$ instances and (b) XORSAT instances. The exponent parameters are (a) $0.25496\pm0.00198$ and $0.43019\pm 0.00274$, and (b) $0.04881\pm 4.8567\cdot 10^{-4}$, and $0.03928\pm 6.7108\cdot 10^{-4}$. To make these plots we used the same data as for Fig.~5 (main text).}
\label{fig:4}
\end{figure}